# Research in teaching-learning sequence design: To what extent do designers' theoretical orientations about learning and the nature of science shape design decisions

**Jenaro Guisasola[1], Kristina Zuza[1,2], Jaume Ametller[3] and Paulo Sarriugarte[2,4]**

[1] Donostia Physics Education Research Group (DoPER-STEMER), University of the Basque Country (UPV/EHU), Plaza Europa 1, 20018 San Sebastian, Spain
[2] Department of Applied Physics, School of Engineering Gipuzkoa, University of the Basque Country (UPV/EHU), Plaza Europa 1, 20018 San Sebastian, Spain
[3] Department of Specific Didactics, University of Girona, Girona 17003, Spain
[4] Department of Applied Physics, School of Engineering Bilbao, University of the Basque Country (UPV/EHU), 48013 Bilbao, Spain

## Abstract

Over the last three decades, various didactic proposals have been published in an attempt to connect theory and research findings with the design of Teaching-Learning Sequences (TLS) in various contexts. Many studies have analysed the process of designing teaching-learning sequences as a research activity. This line of research aims to increase the impact and transferability of educational practice.

However, the information usually provided about the relation between the theory and research findings with the design of the TLS is insufficiently detailed to provide the basis for a critique. Furthermore, not all TLS proposals include evaluation in terms of learning outcomes and very rarely are these learning outcomes specifically related to the design process. This lack of detailed information on the design and evaluation of proposed TLS makes it difficult to properly assess their potential effectiveness or to systematically discuss and improve their design. In this chapter we want to contribute to make the rationale for design decisions explicit. The aim of this paper is to describe in detail how the theoretical orientations of designers of teaching materials towards cognition and learning can shape the structure and pedagogical strategies of the resulting TLS. We will analyse the relationship of two design tools (Epistemological analysis and Learning demands) to theoretical assumptions about learning and the nature of science. We want to highlight the benefits of reflecting on and discussing theoretical elements and their links to design decisions, which makes TLS design more productive on a practical level to broaden the teaching and learning perspectives of TLS. Finally, we will explore the question to what extent the theoretical orientations of curriculum designers towards cognition and learning can influence the structure and pedagogical strategies of the resulting TLS

## 1. Introduction

Since the second half of the 20th century, research into physics teaching has emphasised the importance of comprehending what students already know and understand [1]. In general, it is accepted that 'prior knowledge can interfere with or facilitate new learning' [2]. A large quantity of research over various countries and time periods has shown that students might have ideas about scientific topics that differ from and are often inconsistent with the canonical principles and theories [3]. The research is described using a range of terms such as misconceptions, intuitive theories and alternative conceptual frameworks. However, there is no wide consensus on the meaning of these terms and there is frequently little coherence regarding how they are used in the literature [4]. In this chapter, we will use the term 'students' conceptions', many of which (when not consistent with canonical science) are considered 'alternative conceptions. The literature also indicates that detected conceptions might vary according to the degree of comprehensibility, plausibility and usefulness by which they are conceived by whoever detected them and that this diversely influences students' learning of canonical scientific ideas [5].

Students' alternative conceptions have been shown to resist traditional teaching. It is only rarely that well-organised lectures given orally to students manage to transform students' alternative conceptions into more scientific conceptions [6,7]. Since the 1980s, various teaching approaches have been proposed that attempt to change students' alternative conceptions into scientific ones [8,9]. Teaching by means of conceptual change proposed in 1987, by Driven et al in the Children's Learning in Science-CLIS [10] project, was subsequently followed by other research teams that introduced some changes and contextualisation in different countries. However, the results were not as good as expected. Occasionally, the 'conceptual change' took place in a minority of students, or it was temporary, and the students returned to their pre-'treatment' conceptions. These outcomes were particularly clear when the students were asked about the same concepts in contexts which differed from any analysed in class [11,12].

The aforementioned criticism brought changes to the focus on 'conceptual change', moving towards interpretations of the students' conceptions that not only considered the conceptual change but also the methodological and epistemological requirements represented by understanding a scientific model [13] This partly implies that students' conceptions can be analysed from different perspectives. The literature shows us a wide range of interpretations from the students which included alternative reasoning patterns that lead to the apparent downfall of many students when it comes to understanding the principles of physics [14] interpretations that analyse the students' conceptions from the knowledge-in-pieces cognitive framework [15] or analysis of conceptions from the framework theory approach which suggests that students usually build 'intermediate frameworks' because they use unscientific epistemic reasoning when assimilating new information which is incompatible with the existing knowledge structure [16]

This diversity in the analysis of the student conceptions ("Alternative ideas", "Knowledge-in-pieces", "Intermediate frameworks") can, at least partly, lead to structuring the teaching-learning sequences considerably differently [17]. Diversity of teaching foci and TLS structures does not represent a disadvantage. The problem is not the diversity of teaching strategies in the materials, but explaining them properly to teachers, so that they can understand any specific justifications that leads to proposing a certain type of material, and make the necessary adaptations without the proposal losing the proven elements on which it is based and on which its claim to effectiveness rests. In this respect, the literature shows that TLS designs frequently lack details regarding how the choice of theory underpins teaching and its design. This lack of detailed information on the design decisions hinders the appropriate evaluation of its potential efficacy or the debate and the systematic improvement of its design. The literature has demonstrated that there are important gaps in the design of curriculum materials. In particular, many papers on TLS design lack: (a) a detailed explanation of the implicit and explicit decisions taken regarding design and implementation; (b) a detailed explanation of the teaching strategies, that are often implicitly processed under the label of 'active teaching' or 'active learning'; (c) a broad assessment procedure (in other words, one that goes beyond the learning that has been achieved); and (d) a detailed description of the iterative refining process. The lack of such explicit descriptions makes it hard for the scientific education community to interpret the results which are presented, propose systemic improvements to the design and base it on findings [18,19]. We argue that the TLS design must be developed even further through the explicit articulation of the methodology, that includes theoretical commitments regarding the research and how these lead to methods for the design, implementation and assessment.

In this theoretical document, we are offering a detailed description, with examples from our research, on how the interpretation of students' conceptions and the conception of the nature of science can be tracked in the teaching structures and strategies that we adopt in our TLS design. This paper does not set out to state that one focus or theoretical framework is superior to others. The principal aim of this paper is to describe in detail how the theoretical orientations of designers of teaching materials towards cognition and learning can shape the structure and pedagogical strategies of the resulting TLS. We will analyse the relationship of two design tools (Epistemological analysis and Learning demands) to theoretical assumptions about learning and the nature of science. We would like to emphasise the benefits of thinking about and discussing the theoretical elements and their ties with the design decisions, that make the TLS design more productive to broaden TLS teaching and learning perspectives.

## 2. The Design Based Research methodology for developing Teaching-Learning Sequences

The Design-Based Research (DBR) methodology attempts to overcome design weaknesses in curriculum materials as described in the previous section, with the aim of not only empirically adjusting 'what works' in a TLS but also developing classroom intervention theories. DBR is a research methodology to generate and prove general teaching-learning theories [20]. The DBR methodology can be defined in several ways, although most authors agree that a DBR project should be developed through cycles of design, implementation, analysis and redesign [21]. This methodology does not imply the use any specific educational theory or specific tools for any of its phases, thereby it affords the researchers considerable freedom on how to implement it. In our approach we propose that the theoretical foundations should be made explicit throughout the three phases in which we specify the design and evaluation of the TLS: a) design; b) implementation; c) evaluation and redesign. In this chapter we will refer to the first phase, which is related to our objective of making explicit the relationship between how students' conceptions are interpreted and TLS design decisions. In this chapter, we are going to refer to the first two phases that are related to our goal of explaining the relationship between the interpretations of students' conceptions and TLS design decisions. We are going to describe in detail how theoretical choices influence the concepts and models which are investigated in the student conceptions and the interpretation that helps the TLS design.

In our work on TLS design and evaluation we propose that first of all, the theoretical foundations guiding the research should be made explicit in three phases: a) design; b) implementation; c) evaluation and redesign. In this chapter we will refer to the first phase which is related to our aim of making explicit the relationship between how students' conceptions are interpreted and TLS design decisions.

### 2.1. The theoretical elements that guide the research

The literature shows that most teaching material designs concur that the understanding and analysis of the scientific epistemology of the specific topic to be taught is a necessary condition at an initial level of the design of any TLS and that any proposal failing to consider how scientists produce knowledge runs the risk of producing students that do not recognise scientific conceptions as rational. Clarification of the scientific topic includes key scientific concepts and principles, points of view on the nature of science and the educational context [22, 23]. Another general consensus on material design is to consider learning from the social-constructivist theory that considers students' conceptions and the social dimension of the school learning [24]. However, the way in which this general theory is used when developing teaching strategies varies from one proposal to another.

In our research to support the scientific learning of specific topics on the school curriculum, we consider the epistemology of science, and we adopt a social-constructivist learning perspective identifying the forms of written and oral expression as mediating instruments between the social and personal plane [25]. This theory of cognition is complemented by a learning theory developed by Vosniadou and colleagues [16, 26] called the 'framework theory' approach. The framework theory approach states that the concepts are included in specific 'framework theories' for the area, that represent various explanatory frameworks of sciences which are currently accepted. Students often construct 'intermediate or hybrid frames' because they use unscientific epistemic reasoning when they attempt to assimilate new information that is incompatible with the existing knowledge structure. Although Framework Theory focuses on the cognitive aspects, we also include motivational and sociocultural aspects.

Our research generally considers the indications given by the new curricula that emphasise integration of concepts and scientific practice. The new standards concur that this is not a case of 'learning the content through a teaching methodology' but that managing to make students see scientific conceptions as more attractive and useful than the spontaneous conceptions requires reiterated and lasting opportunities to implement procedures and acceptance criteria, which are characteristic of scientific work. We understand that the construction of knowledge is not approached as questioning students' ideas but as the result of a process of posing problems whose resolution, guided by the teacher, requires students to integrate new knowledge or reformulate existing knowledge [27, 28, 29]. As we will see in the design phase, this understanding of construction of knowledge influence the design tools at a 'fine grain' level. We use the term 'design tool' in the same way as Ametller, Leach and Scott [30], to highlight that the theoretical and empirical knowledge is used explicitly and intentionally when making design decisions.

## 2.2. The design phase: analysis of the student conceptions and their influence on the design

In the DBR methodology, the design phase explicitly connects the theoretical assumptions with the TLS design. This phase leads to an initial product (the TLS) that includes a hypothetical learning path which is consistent with the general theories. As the DBR is developed in real teaching contexts, one initial task is to explicitly identify contextual elements such as the educational level, the study plan, students previous schooling, teachers' training.... that limit the reach of the TLS. This initial phase also includes the epistemological analysis that provides key concepts, reasoning patterns and scientific models to give a well-substantiated definition of the learning objectives that should be expected from the teaching on this topic. This allows us to explicitly define the learning objectives for the education level targeted in the study.

The explicit application of social-constructivist theory that we have adopted is partly to be found in the definition of "learning demand" as a design tool. The Learning demand [30] design tool is used to analyse the ontological and epistemic differences between the students' ideas and the defined learning objectives. If there are no previous experimental results appropriate to the subject and educational level, specific experimental designs are carried out to sound out the students' knowledge and reasoning in relation to the defined learning objectives. The learning demand tool makes a qualitative evaluation of the differences between the students' ontological and epistemological comprehension on the concepts that are going to be taught and the scientific comprehension planned by the end of the teaching. These differences will guide the TLS teaching trajectory in the presentation of content, highlighting the type and degree of difficulty we might expect learners to encounter. The result of applying this design tool is the identification of the most "difficult" and the "key" points that will need to be emphasised in a more "intense" or "differential" way. In our case it also guides the proposal of "driving problems" and indicates where more effort should be devoted. The learning objectives will be reformulated at this point, if necessary. The initial task of clearly and explicitly defining the learning objectives is crucial if we want TLS assessment results to be useful in future designs.

Once the learning objectives, the possible learning demands, and the learning path for the topic have been determined and justified, an initial version of the activities will be defined, comprising the TLS material to be implemented. This will generate the necessary documents for the implementation, assessment guidelines and material for the teaching staff with information on how to use the work materials (see table 1).

Table 1. Design Phase

| Analysis of the educational context and Epistemological analysis | Analysis of the student conceptions and conceptual and reasoning difficulties | Need for interactive environments that reflect the skills and attitudes of the scientific research |
|---|---|---|
| *Learning objectives* | *Learning Difficulties and Learning Demands* | *Teaching strategies* |
| Constructing specific tasks that lead to a teaching path proposal. *TLS activities. Teachers' guide to implement TLS* | | |

## 3. Design of TLS for the topic of Capacitance and Capacitors in Introductory Physics courses

According to the theoretical elements and the specifics of the design that we defined in the previous section, we will outline the educational context where the TLS will be implemented.

The TLS on 'Capacitance and capacitors' was designed on a transformed calculus-based physics course for first year students of engineering and sciences at the University of the Basque Country (UPV/EHU). Introductory Electromagnetism is taught at the UPV/EHU during the spring term (14 weeks) to several groups of 60 students. The traditional course format is 2 hours a week of lectures and 1.5 hours a week of problem-solving sessions. The Electromagnetism course teaches the topic of 'Electrostatic Energy and Capacitance' for 2 weeks. The lectures and the problem-solving sessions cover electrostatic potential energy, capacitance and capacitors, batteries and circuits, energy stored in capacitors and dielectrics. The teaching also looks at analysis of charging and discharging a capacitor. Problems are tackled which are similar to those found at the end of the chapter [31].

On traditional courses, the students do not normally have the chance to actively take part and they are limited to taking notes on the teacher's explanations, both in lectures and in problem-solving sessions. The transformed version follows the same study plan (in other words, covers the same factual knowledge in the same time interval). However, as explained, the course and the content are organised differently. Students that take the introductory physics courses at the UPV/EHU have three semesters of Physics under their belts at High School (16-18 years old) (mechanics, electromagnetism and modern physics) and have passed the University entrance exam. This implies that the students already have basic knowledge regarding electrostatics and DC-circuits. Students have a similar background and their random distribution into five Introductory Physics groups guarantees that student knowledge is reasonably uniform in all groups.

Below, our assumption of framework theory approach implies investigating the students' different 'explanatory categories' on the specific topic. The term 'explanatory category' refers to the model that students use to interpret the phenomena studied and the epistemology they use to understand them [32]. In other words, students cannot be expected to assimilate the conceptual contents if the procedural and ontological aspects are not considered [33]. Investigating the students' different intermediate models means we must ask the students about relevant aspects or 'conceptual keys' for the topic to be learnt. These conceptual keys partly emerge from knowledge of epistemological and ontological problems that had to be overcome to build the theories included in the curriculum [34]. A close look at the students' answers, when they analyse questions that include the topic's conceptual keys, gives us information on the students' conceptions and how we might categorise them. These categories are not understood as 'conceptual errors' but as 'intermediate states' with conceptual and reasoning elements that are particularly difficult in the learning.

### 3.1. - 'Epistemological analysis' design tool

To investigate the students' conceptions on the topic, we need to define the key aspects of the scientific model. In our research, we use the 'Epistemological analysis' design tool that epistemologically justifies the teaching goals in the topic for the chosen education level. Let's start by analysing the epistemological development of the concept of electrical capacitance and capacitors.

During the 18th century, the concept of electrical capacitance became a key concept to explain the processes of charging and discharging capacitors and insulated conductors (Leyden jar). In the early 18th century, electrical charge was considered to be a fluid or substance and the charged body was the recipient where it is found. Interaction between charged bodies takes place due to the contact between its 'electrical atmospheres' [35, 36]. Consequently, the capacitance of a body is understood as a property similar to the volume of the 'recipient body' giving the idea of charges that can be stored. This implies that the larger the surface of the body, the greater the charge will be and that the presence of electrified bodies very close together does not influence their charging process. Subsequently, attempts by Franklin, Volta and Cavendish to satisfactorily explain how the Leyden jar works represented an important qualitative step in electrical theory in the 18th century. It is considered that the electric fluid accumulated in the body exerts pressure on the surface of the body. At a certain point, the 'electric pressure' is great enough to stop the body admitting more charge. Within this theoretical framework, the capacitance of a body is defined as C = Q/Voltage, in a similar way to the historical interpretations by Volta and Cavendish [37]. However, its 18th-century definition is a little ambiguous in reference to the current concepts of 'electric charge' and 'electric potential' [38].

During the 19th century, it was considered that charges can act at a distance. In this respect, the charges accumulated in the body repel each other (Coulomb's Theory) over the surface of the body. Furthermore, the ability of the charges to act at a distance implies that the presence of an electrified body close to the body to be charged can modify its 'electrical voltage' and its 'capacitance' to store charge. This is how the basics of the capacitor and Leyden jar are explained. Scientists came up against the explanation of phenomena related to the process of charge accumulation in bodies and on the electrical nature of matter (conductor-insulator dichotomy). Solving these problems therefore implied the need to define new magnitudes such as electrical potential and electrical capacitance. These concepts evolved until they were used both in electrostatic and electrical current contexts and took on their current meaning in the theoretical framework [39]. The process of charging a body represents work that implies acquisition of an electrical potential. Consequently, the concept of capacitance is a property of the system of conductors that interact and that in principle can be measured as shown in the equation: C = Q/ $\Delta$V, applying a known charge and determining the potential variation. This theoretical framework corresponds to an energy interpretation of the concept of potential [40].

The previous discussion on historical development provides important clues regarding the key ideas behind the concept of capacitance. We have pinpointed four key characteristics of the electrical capacitance concept that we consider relevant and attainable at a university introductory physics level [38, 39, 41, 42]:

**K1.** During the process of charging a body, the energy stored in it can vary. The explanatory model includes the concepts of charge and electrical potential energy.

**K2.** The previous model determines relationships between the concepts of charge and electrical potential that define the concept of *electrical capacitance* for a body. This can be measured macroscopically (by means of current and electrical potential difference) in a circuit.

**K3.** The explanation based on the charge/potential relationship (electrical capacitance model) implies that the presence of other charged bodies around a body to be charged can improve its charging process.

**K4.** Therefore, a system formed by two conductors that are close to each other (with total influence) with opposite charges, in other words a capacitor, optimizes its electrical capacitance. The role of dielectrics in increasing or decreasing the capacitance of a capacitor.

The key ideas identified using the Epistemological Analysis tool leads us to a well-substantiated definition of the learning objectives for the topic at the chosen education level (see table 2).

Table 2. Epistemological justification of the learning objectives on electric capacitance and RC circuits

| Epistemology of physics issue | Learning objectives |
| --- | --- |
| The electrical capacitance model was determined to explain the processes of charging a body (K1 and K2). | O1.- To explain that when a body is charged, it acquires energy due to the work done by the environment-battery to charge it. <br><br> O2.-To understand the concept of electrical capacitance as the relationship between charge and electrical potential. |
| The explanation based on the charge/potential relationship (electrical capacitance model) shows us that a system formed by two conductors that are close to each other (with total influence) with opposite charges, in other words a *capacitor*, optimises its electrical capacitance. The role of dielectrics in increasing or decreasing the capacitance of a capacitor (K3 and K4) | O3.- The capacitance of a capacitor is the proportional constant between the amount of charge on the plates and the potential difference across the plates <br><br> O4.- To understand the influence of the dielectric and geometric factors of a system on its electrical capacitance. |

## 3.2. 'Learning demand' design tool

The study of student conceptions is guided by the defined learning objectives. The study does not analyse the students' ideas on potential energy and electrical capacitance in general, but on the electrical capacitance model and its conceptual elements included in the learning objectives. A questionnaire was designed for this with seven questions, inspired by our previous studies [43, 44, 45].

Questions Q1 and Q2 investigate the role of potential difference in the process of charging a body (objective O1). Questions 3 and 4 explore the students' comprehension relating to the concept of electrical capacitance (objectives O1 and O2). Questions 5, 6 and 7 investigate the influence of the environment on the processes of charging the bodies (objectives O3 and O4). We design open-questions for students because these questions require students to use 'creative reasoning' based on the scientific content of the task rather than 'imitative reasoning' (memorized reasoning, remembering an algorithm and calculating the answer) based on superficial properties of the task [46]. Furthermore, a semi-structured interview was designed that uses a Prediction-Observation-Explanation (POE) format [47] to look at the students' explanations in greater depth. The interview took place on Volta's electrophorus phenomenon.

The questionnaire was validated by three physics teachers , and it was given to a small sample of students of an electromagnetism module in the first year of an engineering degree. Once the corrections suggested by the teachers, and the insights from the analysis of results of the pilot, had been introduced, , the questionnaire was given to 161 first year students of "Electromagnetism", post-instruction. The questionnaire was part of the continuous assessment process. The students filled in the questionnaire under exam conditions (in silence), taking between 30 and 40 minutes. Furthermore, 7 volunteer students were interviewed.

The objectives of analysing the answers were, firstly, to identify a set of description categories for each question and, subsequently, to classify the answers according to these descriptive categories. The analysis was based on a phenomenographic methodology [48]. When categorizing the answers, we coded the comments understood to be 'an explanation' for the occurrence of some easily recognizable features, such as scientific statements and argumentation. To do so, two of the authors identified the categories for a small subset individually; they met, discussed and reviewed the categories, they tried them in a second subset, and then met, discussed and reviewed the categories. This procedure went on until a reasonably stable set of categories had been produced. Once the final descriptions had been agreed on, a final sample of 25 answers was selected and categorised by the three researchers in this study. A high degree of consensus was reached with an average score for Cohen's kappa reliability score of 0.87, which is accepted as a reasonable level of agreement for this type of research. Subsequently, one of the researchers classified the remaining answers according to these categories.

For the sake of brevity, we are only presenting two of the questions, using them to illustrate some of the students' explanatory patterns.

## Question Q2. The process of charging a body

**Q2.** *Two spheres of equal radius R, one of metal and the other of plastic, are connected  separately (see figures) to a 12v  battery. Which of them will acquire more charge? Why? Describe the process of charging*

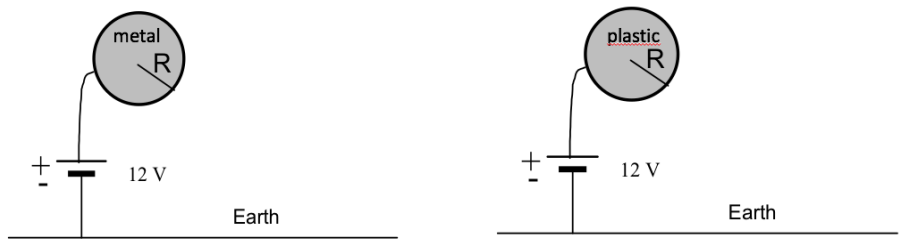

## Question Q6. The concept of capacitance in a capacitor.

In the diagram, conductor $C_1$ has been charged +Q by connecting it to the V volt battery (situation a). The ammeter A indicates there is no current. A neutral body $C_2$ is brought close to it. Conductor $C_2$ is then grounded, and the ammeter shows that a current passes during a brief period of time along the wires connected to conductors $C_1$ and $C_2$ (situation b),
- How can you explain this phenomenon?
- Does the capacitance of conductor $C_1$ change when conductor $C_2$ is brought close to it?

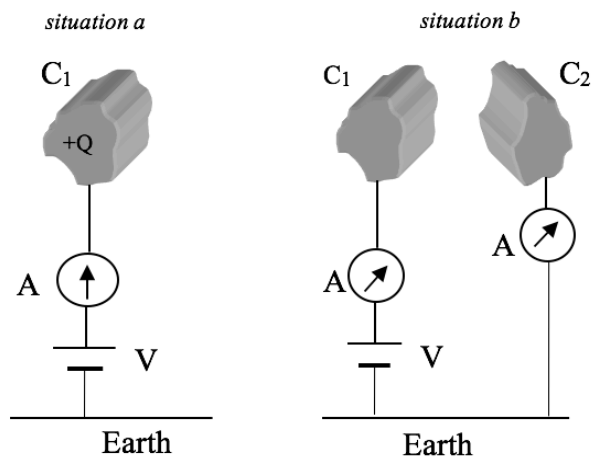

Question Q2 presents the charging of two spheres made of different material (conductor and dielectric). The students must assess the distribution of the charge over the sphere, explaining that the charge in the conductor is spread uniformly over the surface, while in the dielectric, it is mainly in the area where the wire touches the sphere. This makes us consider that for the same work per unit of charge (same voltage), the conducting sphere will accumulate more charge. 13.5% of the answers are correct and reasoned according to the concept of potential difference and the electrical nature of the matter. A student explains it as follows:

*"The conductor will charge very easily at first because there is no repelling charge and as it accumulates on its surface the potential will increase. When the potential of the body reaches that of the battery, the charging process will end"*

In question Q6, the students must base their reasoning on the system's potential difference in situation a) and in situation b), explaining that the proximity of the conductor $C_2$ polarises the conductor $C_1$ and there is a drop in the potential difference for the system $C_1$- $C_2$ which makes current flow from the battery to the conductor $C_1$. In addition, the polarisation of the conductor $C_2$ means that there is a flow of current in the cable connected to earth. 7.5% of the answers explain the phenomenon properly. For example:

*"If there is charge movement in situation b), this is because there is a difference in potential between the battery and the body. In situation b) I have assumed that the potential of body $C_1$ may, in some way, perhaps through electrostatic influence, vary body $C_2$. There is an electrical potential variation in the system formed by both bodies and therefore a movement of charges in situation b). Varying the charge and potential difference will vary the electrical capacitance of the system C1 - C2"*

The vast majority of explanations use elements from the theoretical framework of physics although with reasoning based on models that are not scientific. We have grouped together the explanatory models that are used in all the questions in table 3.

**Table 3.** The students' explanatory categories

| Explanatory categories | Charging process | The role of potential difference | The environment and close bodies | Capacitance |
|---|---|---|---|---|
| amount of charge | difference between the amount of charges | No explanation | No explanation | confusion between capacitance and amount of charge |
| Coulombian force | Charges moved by Colombian forces | confusion between force and potential difference | Electrical interaction by coulombian forces | Description of the equation $C=Q/\Delta V$ |
| Based on the equation | No explanation | No explanation | No explanation | Description of the equation $C=Q/\Delta V$ |
| Scientific model | relationships between charge and electrical potential | Electrical energy is Fundamental in the charging process | Changes due to variation of potential difference | Capacitance as the relation between potential difference and the amount of charge |

We found that around 85% of the answers were included in alternative rather than scientific explanatory categories. The 'quantity of charge' category is characterised by explaining the charging processes of a body, by means of charges passing due to the difference in the quantity of charge between the connected bodies (39.5% in Q2 and 45.5% in Q6). One student explained:

*"The metal sphere will be charged until the charge in the battery and the sphere is the same. There is a transfer of electrons from the most charged body (the battery) to the least charged (the metal sphere)*

*until the quantity of charge is balanced. The plastic is insulating and as it does not conduct electricity, it does not let charge pass through. The plastic sphere does not charge" (question Q2).*

In this explanatory category focused on the magnitude of electrical charge, the explanations only consider the body to be charged, ignoring the role of the environment. Another student explained:

*"Charging will be easier when another conductor is brought closer, the spheric crust, as long as the conductor being brought closer has a higher charge. Then conductor $C_2$ will tend to charge $C_1$ and the quantities of charge will tend to equal out. It is easier to charge two coupled bodies (a capacitor) than just one. As it has more charge, the body $C_1$ has more electrical capacitance" (question Q6)*

The 'Coulombian force' category includes explanations that analyse the charging process considering the electrical charge magnitude and the action at a distance with other charged bodies nearby. In this category, the influence over the environment depends on electrostatic interaction. This vision leads to identifying the potential of the system with the Coulombian force that the system's bodies exert (12.5% in question Q2 and 15% in question Q6). The attracting forces increase the system potential, and the repelling forces reduce it. This explanatory model does not consider medium polarisation phenomena. For example:

*"The battery performs the electrical force to bring the charges to the body. The body exerts a repulsive force as it becomes charged until it acquires charges" (question Q2)*

*"By bringing body $C_2$ closer, electrical forces are exerted, meaning that body $C_1$ can receive more charge and increase its electrical potential" (question Q6)*

The 'based on the equation' category includes explanations that use the description of the formula for electrical capacitance C=Q/V as their only argument, although without managing to explain the meaning (31.5% question Q6). As they do not find meaning in the concepts, the students 'take refuge' in the operative definitions and reason from them. One student explained:

*"A body's capacitance is C=Q/ΔV. In situation b), the system is a capacitor (lowest ΔV) and therefore it has greater capacitance that the body $C_1$"* (question Q6)

The description in explanatory categories helps reveal learning difficulties that lead to 'intermediate explanatory categories' that lie in a learning process reasonably close to the scientific explanation [16]. We have defined these difficulties and used the 'Learning demands' design tool. This design tool is based on our social-constructivist assumption of the learning, particularly in the student conceptions concepts and the 'zone of proximal development' (ZPD) [49]. The ZPD is defined by the space between what the students can do for themselves and what they can do with help from an expert. Therefore, the ZPD can be considered as the distance between what the students know and what they need to learn, provided they are capable of achieving that goal with the assistance of the teacher. The distance is not a direct measure of the difficulty of the learning objective, but it is clearly related to it. The 'learning demand' tool attempts to evaluate the ontological and epistemological distances between the student conceptions and what we have picked as learning objectives, in the specific TLS that we are designing. When it comes to defining the Learning Demands, we consider three aspects:

i) Degree of Inconsistency with Scientific Models: one way in which the student conceptions vary is the degree that separates them from the canonical knowledge, in other words, to what extent their conceptions differ from the scientific concepts.

ii) Degree of Connectedness of Student Knowledge: Some of the conceptions can be fragments of knowledge that are reasonably isolated, small 'islands of knowledge'.

However, other conceptions are firmly connected in networks that can be mutually strengthened.

iii) Degree of Commonality of Student Knowledge: the frequency with which the conception appears in the literature.

In the specific case that we are analysing, table 4 compiles the defined difficulties and learning demand.

**Table 4.** Learning objectives, difficulties and Learning demand

| Learning Objective | Students' difficulties on key scientific ideas | Learning demand |
|---|---|---|
| O1. Process of charging | D1.-Students focus on the amount of charge or electric foce to explain the charging process of a body. They do not consider the concept of potential difference of the system to explain its charging process. | Conceptual; High |
| O2. Capacitance | D2.- Students do not correctly understand the concept of electrical capacitance as the relationship between the charge and the electrical potential it acquires in the charging process. | Conceptual and epistemological; High |
| O3. Capacitor | They cannot explain the electrical influence between bodies and the role of the dielectric in a capacitor. | Conceptual (behaviour of matter under electrical interaction); High |
| O4. Enviroment influence | D3.- Students use a reasoning based on the equation. Consequently, only take into account the size of the objects and the quantity of charge that they have. | Epistimological (based on the formula). Medium |

### 3.3. How theoretical assumptions and design tools shape the structure of TLS: Teaching approach focused on Learning by Guided Problem Solving (GPS)

Theoretical assumptions about learning and about the epistemology of science lead us to pay attention to the key ideas of the topic and to the students' initial conceptions about them and how to overcome them. The defined didactical tools provide us with the necessary information. TLS structure organisation starts by identifying problems that are found at the root of the scientific models the students must learn. We propose a structure of activities based on a sequence of 'driving problems' that the students are set to help them get a preliminary idea of the tasks that they must perform to attain the learning objectives. The driving problems chosen to guide the students' learning on the topic of 'Capacitance and capacitors' are shown below:

- What is the interest of storing charge in bodies?
- How is charge stored in a body?
- What is the electrical capacitance of a body?
- What systems allow us to store large amounts of charge?
- Electrical capacitance of a capacitor. How is it possible to increase the capacitance of a capacitor?

For each 'driving problem', a set of activities is designed that provides opportunities for the students to appropriate epistemic reasoning that helps them progress towards a scientific framework. As an outcome, we design a course structure that allows the students, guided by the teacher, to address problems of interest, putting into play a large number of the processes to produce and validate scientific knowledge [50] (Becerra et al 2012). The designed activities give the students opportunities to practice production and validation procedures on the scientific knowledge (analysing problems, making hypotheses, arguing based on evidence, analysing results, etc.). We call this teaching focus 'Teaching by Guided Problem Solving'.

Examples of activities designed as scaffolding to set, analyse and solve the 'driving problems' are given below. Firstly, the first two activities for the problem: How is charge stored in a body?

A.1. A battery with a potential difference $V_0$ is connected to a metal sphere (see diagram). Will the sphere be charged? Explain how the current flows from the battery to the sphere. What is the final electrical potential of the sphere?
If the sphere was made of plastic, would the charging process and the final potential of the plastic sphere vary?

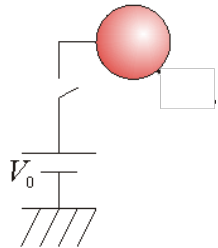

A.2. The figure shows a conductor with zero net charge and insulated (situation a). When connected to a 12 V battery, current flows through the ammeter (situation b). After a short time, the current stops and the sphere has a positive charge Q (situation c).
- Why do electrons circulate in situation b? (Note: remember how negative charges move in an electric field).
- Why does the current of electrons stop?
- What is the electric potential of the sphere in situation c?

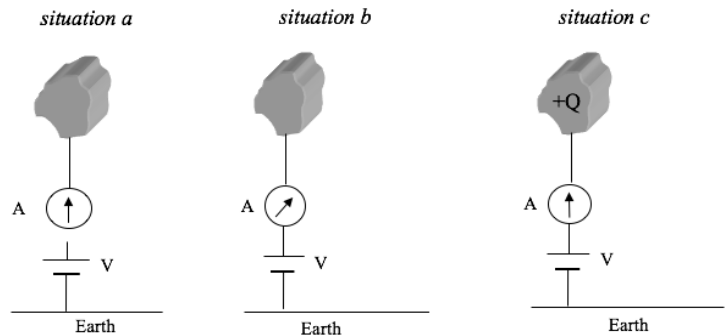

The activities ask explicitly about the charging process and the magnitude of the electrical potential that is frequently not used by the students (difficulty D.1.). This refers to analysing the charging process taking into account that, as studied in electrostatics, the charge must flow along the cable until the potential is equal in the battery terminal and the conductor. It is important to analyse the process with different materials (conductors and dielectric) and see the influence of the distribution of charges and the electric potential of each sphere in activity A.1.

In relation to the problem: What is the electrical capacitance of a body? The first activity is given below

A.1. What potential difference would a battery connected to a conducting sphere of twice the radius of another have to have in order to accumulate the same amount of charge on both?
Which of the two spheres has the greater electrical capacitance? Explain.

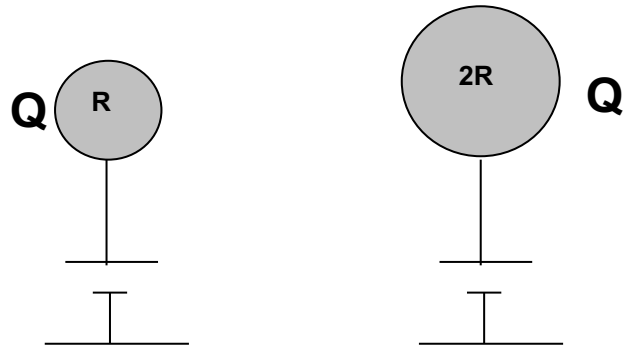

From the start of the process of defining the capacitance concept, the electrical potential concept is explicitly introduced plus its relationship with the work of charging the body. The qualitative (and later quantitative) relationship between quantity of charge and electrical potential would be the starting point for introducing the concept of electrical capacitance.

## 4. Discussion

The analysis and detailed justification of the design decisions in relation to the theoretical orientations, that we explained in the previous section, lead us to state that the designers' theoretical orientation influences the design of the TLS but does not determine it unequivocally. Other influences cannot be disregarded such as pedagogic knowledge of the content and the professional experience of the authors and their research groups, the educational context, the communication customs in class, etc. In fact, the bibliography shows that research teams with similar theoretical orientations and the same data on the students' difficulties do not necessarily agree on the same TLS design [51,52]. The causal connection between the theoretical assumptions of the research group and the TLS design is complex and, as we mentioned, depends on multiple factors. What we argue in this paper is that some differences between TLS design foci might be due to different theoretical orientations from the designers.

In this paper, we have not presented evaluation data for the TLS but the reader can consult prior evaluated implementations on the topic of "Electrical capacitance" [53] and other curriculum topics implemented and assessed from the same theoretical assumptions [54]. However, our goal in this paper is to demonstrate the influence of the research group's theoretical assumptions on the TLS design. In our opinion, the analysis of this problem requires its own space for reflection such as developed in this document.

## 5. Conclusion

'Framework theory' and the epistemology of science have led us to design a TLS structure that brings the students' intermediate models and the learning demand to the fore. The research intended to epistemologically justify the students' learning objectives. It has described the learning demand between these objectives and the student conceptions. To do so, we have investigated the students' explanatory categories relating to these learning objectives. These explanatory categories show us the students' reasoning frameworks, which they apply to different charging phenomenon for bodies and capacitors. From this focus of understanding the student conceptions, at least partly, we can formulate 'driving problems' so that students can propose their intermediate models and refine them towards the scientific model with help from scaffolding activities that guide the students towards solving the problem. In the Guided Problem-Solving pedagogic strategy that we have defined, the use of scientific practice to resolve and interpret phenomena plays a fundamental role for the conceptual and epistemic flow from intermediate to scientific models.

We have attempted to answer the question that we raised at the outset, on how far the curriculum designers' theoretical orientations towards cognition and learning can influence the structure and pedagogic strategies of the resulting TLS. For this reason, we have presented a detailed design for a TLS by explicitly demonstrating the basis of the design decisions in the theoretical assumptions of our research group. We illustrated how theoretical elements help us to define design tools such as 'Epistemological analysis' and 'Learning demand' and that these tools give meaning to the final specifying level in the TLS activities.

The debate demonstrated in this paper does not manage to set a causal relationship between the theoretical elements of the research team and the TLS design, but it does show that some features of the design focus might be due to the theoretical orientations. Analysis and conscious justification of the theoretical orientations underpinning the design of curriculum materials helps us to understand the proposed TLS and provides new ideas for its revision. We uphold that the reflective interpretation of the students' answers from explicit theoretical suppositions can guide the design of the materials and their subsequent refinement. We hope to contribute to the debate that emerged in the PER community on the need to give a substantiated explanation for the design decisions and how to integrate the generating role of the theory and the empirical data.